\documentclass[preprint2]{aastex63}
\usepackage[utf8]{inputenc}
\usepackage{caption}

\usepackage{colortbl}

\shorttitle{Spectral Classification of Algol C}
\shortauthors{Frank, Whelan, \& Junginger}

\begin{document}

\title{Spectral Classification of Algol C}

\author{Megan G. Frank}
\affil{Department of Physics, Austin College, Sherman, TX}

\author{David G. Whelan}
\affil{Department of Physics, Austin College, Sherman, TX}
\email{dwhelan@austincollege.edu}

\author{Jessica C. Junginger}
\affil{Department of Physics, Austin College, Sherman, TX}



\begin{abstract}
The spectral classification of Algol~C, the third star in the Algol triple system, has long been a matter of some uncertainty. There is good reason to suspect that it should be a metallic-line A-type star, and one study in particular showed that this was so, but further studies have cast doubt on that assessment. We utilized a simple spectral subtraction method between spectra taken in and out of primary eclipse to reveal the light of Algol~C in the absence of the light from the brightest star. Our resultant spectrum is well-matched to an F1~V spectroscopic standard and shows no evidence of metallic-line spectral anomalies. We note that this classification matches recent abundance determinations for this source.
\end{abstract}

\section{Introduction}\label{sec:Intro}

Algol ($\beta$ Persei) is a hierarchical triple system in which Algols~A and B are the eponymous eclipsing pair.  Algol B, the less massive secondary, is physically large, filling its Roche lobe and losing mass to the primary. The orbital period of Algols~A and B is a short 2.87~days. The third component of the system, Algol~C, has an orbital period of 680~days around the binary pair \citep{Zavala2010}.

Algol~A is by far the brightest star in the system, contributing $\gtrsim94$\% of the light at visual wavelengths \citep{Kolbas2015}. Algol~C is the next brightest, contributing $\gtrsim5$\% of the system's light. The dimmest component of the system is Algol~B, which contributes $\lesssim0.8$\% of the system's light. During primary eclipse, the majority of Algol~A is covered up by Algol~B. This causes spectral absorption lines from Algol~C to become prominent, allowing for identification of the star's spectral type.

The spectral type for Algol~C seems uncertain from the literature. \citet{Hall1939} claimed it was approximately an A5~V star, while \citet{Meltzer1957} approximated F5~V based on absorption line ratios. Other spectral types \citep[{\em e.g}.,][]{Struve1957} were approximate and while many attempts were made to answer this simple question none could be more precise. \citet{Fletcher1964} determined that Algol C was a metallic-line A-type (Am) star, specifically with a spectral type of kA4hA9.5mF0 for the spectral classes of the Ca~{\sc ii}~K line, hydrogen Balmer lines, and metal lines respectively. The presented evidence for this spectral type included that the Sr~{\sc ii} $\lambda 4077$, $\lambda 4216$, and Ca~{\sc i} $\lambda 4226$ lines were all slightly strong for an A9.5 star, and that the Ca~{\sc ii}~K line was particularly weak, which is common for Am stars. The strengths of these lines relative to the Fe~{\sc i} lines as well as a magnitude difference between Algols~A and C ($\delta M_{ac} = 2.6$) were used as supporting evidence. 

Physical considerations justify an Am designation for Algol~C. Algol~C is a very slowly-rotating star, with a rotational speed of $\approx~12$~km~s$^{-1}$ \citep{Kolbas2015}. Meridional circulation is responsible for chemical mixing in A- and F-type stars that do not have convection zones near the surface, and the star must be rotating at $\gtrsim$~90~km s$^{-1}$ for this effect to take place \citep{Charbonneau1993}. All the Am stars have slower rotation rates than this, though it should be noted that not all slowly-rotating A- and F-type stars are Am stars \citep{Abt1985}. However, nearly all Am stars are part of close binaries in which the Am star is in synchronous rotation \citep{Gray2009}. Algol~C is not in synchronous rotation \citep[using the rotation speed from ][]{Kolbas2015}, so other evidence is needed to confirm that Algol~C is an Am star. 

\citet{Richards1987} determined that the spectral type for Algol~C is F$1\pm1$. They particularly noted that the Ca~{\sc ii}~K line was only marginally weak for the spectral type, contrary to \citeauthor{Fletcher1964}'s findings. \citet{Richards1988} followed up with a temperature for Algol~C of 7500$\pm$500~K, consistent with a late A- or early F-type star. Later, \citet{Richards1993} compared the metal line ratios for Algol~C to four other stars (spectral types A7~V to F5~V) using Mn~{\sc i} $\lambda 4030$, Fe~{\sc i} $\lambda 4046$, Ca~{\sc i} $\lambda 4226$, and Fe~{\sc ii} $\lambda 4233$. They showed that the metal line ratios for Algol~C were consistent with an F1 or F2 spectral type, with the exception of the Calcium line, which was underabundant. However, their analysis of the Balmer line profiles suggested a temperature of 7500~K, which is too hot for an early F-type star but consistent with their earlier study. Their analysis of the Balmer lines therefore suggests an earlier spectral type than does their abundance analysis.

The most recent study of the metal abundances of Algol~C is found in \citet{Kolbas2015}. This work disentangles the spectra of the three components in the frequency domain, so that spectral absorption lines are perfectly preserved for each component's spectrum. They measured the abundances of twenty-two elements for Algol~C and compared them to solar abundances and the abundances of chemically normal and Am-type stars from the Hyades open cluster study of \citet{Gebran2010}. They show that Algol~C has a roughly solar abundance for all elements measured, and of particular note is the fact that they do {\em not} see any underabundance of Calcium in its spectrum.

We present a spectrum of Algol~C that we then use for spectral classification. Section \ref{sec:Obs} discusses the observations and data reduction, Section \ref{sec:Method} examines the methods used to isolate the spectrum of Algol~C, Section \ref{sec:Analysis} presents the analysis of our final spectrum, and we conclude in Section \ref{sec:Conclusion}.  

\section{Observations and Data Reduction}\label{sec:Obs}

\begin{deluxetable*}{ccccc}[hbt!]
    \tablenum{1}
    \tablecaption{Observational Log}\label{tab:obs}
    \tablehead{
        \colhead{Date} & \colhead{Time} & \colhead{No. of} & \colhead{Sec Z} & \colhead{Phase}\\
        \nocolhead{} & \colhead{Range} & \colhead{Spectra} & \colhead{Range} & \colhead{Range}
    }
    \startdata
    2020-08-14 & 08:32-09:08 & 3 & 1.344-1.223 & 0.31-0.32\\
    2020-08-18 & 08:11-10:32 & 12 & 1.366-1.044 & 0.70-0.73\\
    2020-08-20 & 08:13-10:06 & 12 & 1.326-1.063 & 0.40-0.43\\
    2020-09-08 & 05:34-09:39 & 86 & 1.812-1.021 & 0.99-0.04\\
    2020-09-18 & 06:21-08:01 & 14 & 1.317-1.078 & 0.49-0.51\\
    2020-09-25 & 04:15-04:44 & 6 & 1.920-1.678 & 0.90\\ 
    2020-09-26 & 06:57-08:15 & 10 & 1.131-1.030 & 0.28-0.30\\
    2020-09-27 & 04:16-04:58 & 6 & 1.838-1.540 & 0.59-0.60\\
    2020-10-06 & 07:07-08:57 & 9 & 1.057-1.012 & 0.77-0.80\\
    2020-10-07 & 04:17-06:07 & 11 & 1.547-1.146 & 0.08-0.11\\
    2020-10-13 & 05:04-06:15 & 10 & 1.247-1.088 & 0.19-0.20\\
    \enddata
\end{deluxetable*}

Low-resolution spectra were obtained using a Lhires III long-slit spectrograph, built by Shelyak, on the Adams Observatory 24-inch $f/8$\ telescope at Austin College. The CCD used to capture the spectra is an e2v $42-10$\ with $13.5~\mu$m pixels. The spectrograph is in the Littrow configuration with a 35 $\mu$m slit, which measures $1.5$'' on the sky. We used a $1200$~gr/mm grating for a dispersion of $0.54$''/pixel. The resolution of the spectrograph is therefore $1.4\mathrm{\AA}$. Observations were taken between 2020-08-14 and 2020-10-13; a log of observations is shown in Table~\ref{tab:obs} with dates, times, number of spectra secured per observing session, the range in airmass, and the range of orbital phase of the Algol AB system at the time of the observations. Observation dates and times were chosen to observe the system at every tenth of the Algol AB orbital phase. Exposure time was $30$~seconds per image.

Data were reduced using standard procedures written in Python. Science images were bias- and dark-subtracted, and flatfield corrected using images taken with an integrated flat-field lamp. Each spectrum was extracted by first fitting a Gaussian curve to the dispersion profile to find its location and width. Pixel values within two standard deviations of the mean were summed, and sky emission was subtracted based on the average values of the sky on either side of the extraction window at each wavelength. The signal-to-noise ratio was computed at each wavelength using the CCD equation. Wavelength calibration was performed by manually identifying the positions of five absorption lines of known wavelength in air in the extracted spectrum and fitting the correlation between pixel number and wavelength value with a quadratic function. Rectified spectra were created by dividing by the continuum that was calculated by fitting a cubic spline between hand-picked continuum points.

\section{Method of Spectral Subtraction}\label{sec:Method}

\subsection{Method Development}

Spectral disentanglement in Fourier space has become a common method for isolating components of complex spectra \citep[e.g.,][]{Hadrava1995}. But this method requires high spectral resolution, so that individual spectral absorption lines can be resolved and their components can be separated. Our resolution of $1.4\mathrm{\AA}$ is far too large to resolve lines from Algol A \& C, whose radial velocity difference is never more than $77$ km/s \citep{Kolbas2015}.

\begin{figure*}[]
    \centering
    \includegraphics[width = 0.99\textwidth]{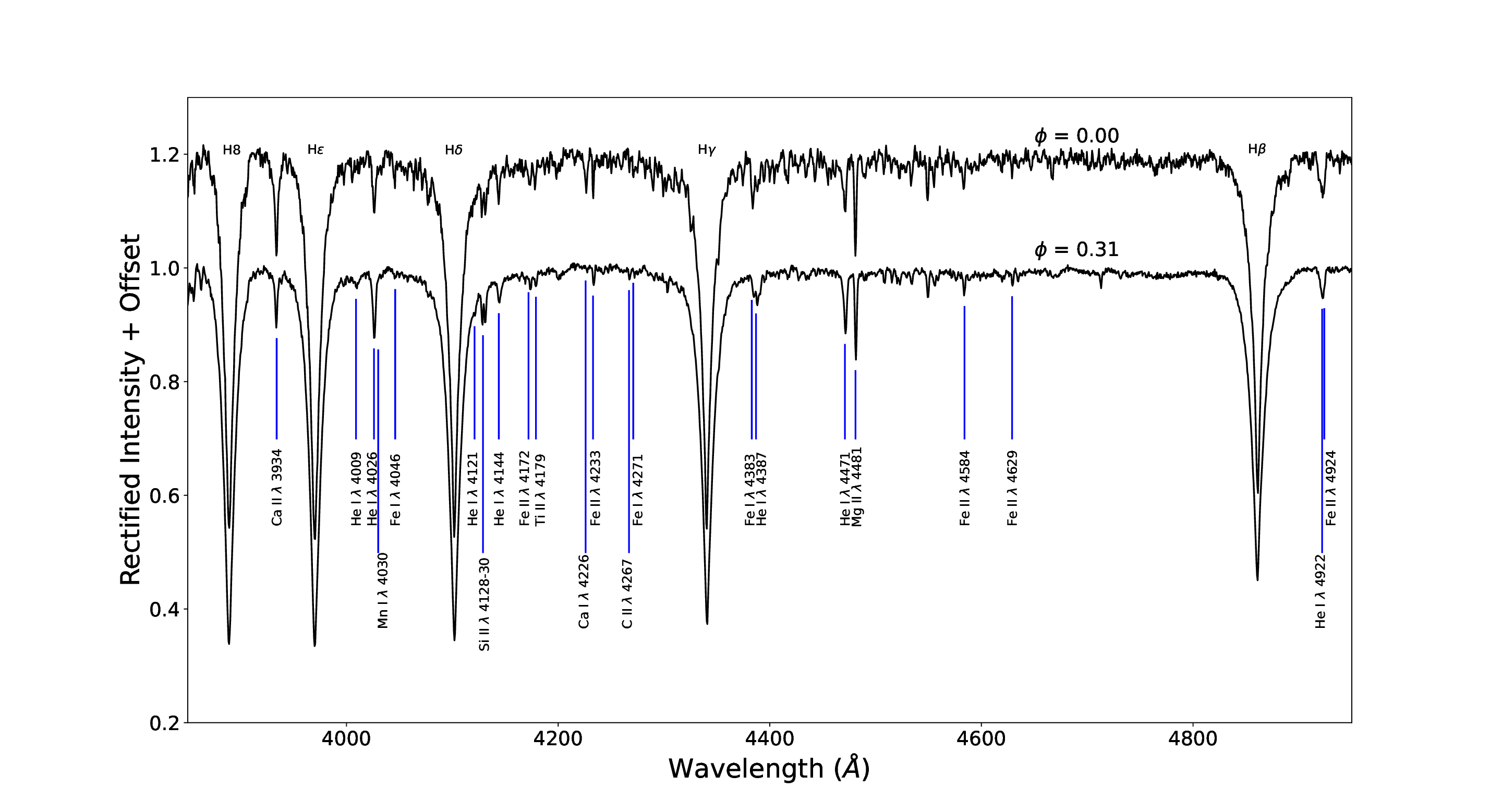}
    \caption{Rectified spectra of Algol taken during primary eclipse ($\phi = 0.00$) and at a phase angle out of eclipse ($\phi = 0.31$), offset by 0.2 units for clarity. Numerous absorption lines useful in spectrally classifying B-, A-, and F-type stars are labeled.}
    \label{fig:Algol_phi01_phi10}
\end{figure*}

A method for determining the spectral classification of companion stars in binary systems using low-resolution spectra was described in \citet{Griffin1986}. One component of the binary is spectroscopically classified, and then the spectrum of a standard star of that class is subtracted from the spectrum of the binary. This should leave only light from the other component, which can then be classified in its own right.

The trouble is that Algol was long considered a B8~V spectroscopic standard in its own right  \citep[e.g.,][]{Cannon1918,Morgan1943}, and is still used that way on occasion today \citep[{\em e.g}., in the library of standard spectra for the expert classifying program MKCLASS;][]{Gray2014}. The most commonly used B8~V standard in the northern hemisphere today is a high rotation-rate star, 18~Tau \citep{Garrison1994,Gray2009}, whose absorption lines are significantly wider than are Algol's. This would present serious issues when subtracting the two spectra. The other accepted standard, HR~9050, has narrower lines but is only visible from more southern latitudes. Therefore we do not have a suitable standard star with which to perform a spectral subtraction.

Our solution is to subtract a scaled spectrum of Algol taken when it is out of eclipse from a spectrum of Algol when it is in primary eclipse. This is tantamount to removing the light from Algol~A, leaving only the light from Algols~B\ \&~C, because Algol~A is by far the brightest star in the system (see \S~\ref{sec:Intro}; what little light from Algol~B that will remain in this final spectrum should be negligible because it is so dim). Representative rectified spectra of Algol taken in and out of eclipse are shown in Figure~\ref{fig:Algol_phi01_phi10}, labeled with the orbital phases at which they were observed. Several differences between the two spectra reveal tantalizing facts about the spectrum of Algol~C. These include the stronger Ca~{\sc ii}~K line, the weaker He~{\sc i} $\lambda 4026$ line, the stronger metal lines throughout the spectrum, and the change in the ratio of the He~{\sc i} $\lambda 4471$ and Mg~{\sc ii} $\lambda 4481$ lines.

\begin{figure*}[]
    \centering
    \includegraphics[width = 0.99\textwidth]{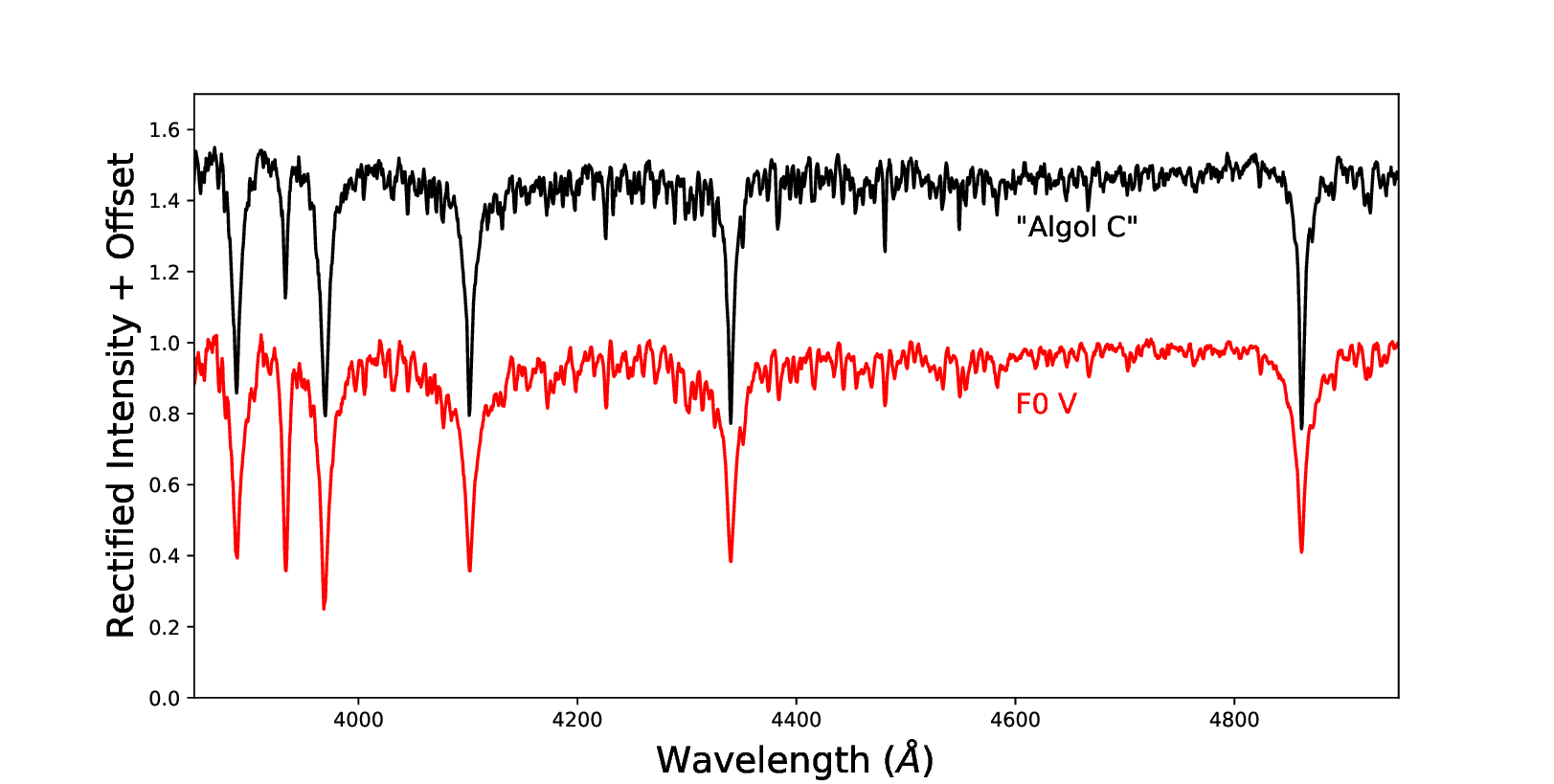}
    \caption{This spectrum of ``Algol C'' (top) was created by subtracting the rectified spectra shown in Figure~\ref{fig:Algol_phi01_phi10}. The standard star spectrum is that of F0V spectral standard HD 23585 (bottom).}
    \label{fig:AlgolC_rect_vs_F0V}
\end{figure*}

Subtracting a scaled spectrum of Algol~A from the spectrum taken during primary eclipse requires that we determine the scale factor that correctly represents the contribution of Algol~A to the total light of the spectrum. We will show some algebra that suggests the range of possible scale values, and we will also show the effect of choosing different scale factors on the final spectrum.

But first we must confront the major issue of how to treat the spectral continuum. Figure~\ref{fig:AlgolC_rect_vs_F0V} illustrates what happens when we subtract continuum-rectified spectra. The subtracted spectra were those shown in Figure~\ref{fig:Algol_phi01_phi10} and they were scaled appropriately before they were subtracted. This procedure should have resulted in a spectrum that effectively only included light from Algol~C. When we compare the subtracted spectrum of ``Algol C'' to an F0V spectral standard, however, we see that the strengths of the metal lines in the Algol spectrum vary with wavelength, when compared with the standard, appearing weakest at shorter wavelengths. This makes it appear as if ``Algol~C'' has a spectral type that is earlier than F0 at shorter wavelengths, but later than F0 at longer wavelengths!

The ultimate reason for the line strength's apparent function with wavelength is that the slopes of the two spectra are different. During primary eclipse a majority of Algol~A's light is being blocked by Algol~B, greatly reducing the contribution of Algol~A to the overall spectrum. This not only affects the strengths of the lines in the spectrum taken during primary eclipse (as illustrated in Figure~\ref{fig:Algol_phi01_phi10}) but also the shape of the continuum, because the cooler stars now contribute more to the overall light of the spectrum.

\subsection{The Finalized Method}

We must therefore subtract un-rectified spectra. An extracted, background-subtracted but otherwise uncalibrated spectrum possesses a continuum that is determined by the starlight extincted by the Earth's atmosphere, and altered according to the spectral response of our instrumentation. Without detailed knowledge concerning the atmospheric extinction or the spectral response function, these factors cannot be removed accurately. But observing Algol in and out of eclipse at similar airmass allows us to assume that the atmospheric extinction is constant. We must also assume that the spectral response is the same for each spectrum.

\begin{figure}[]
    \centering
    \includegraphics[width = 0.49\textwidth]{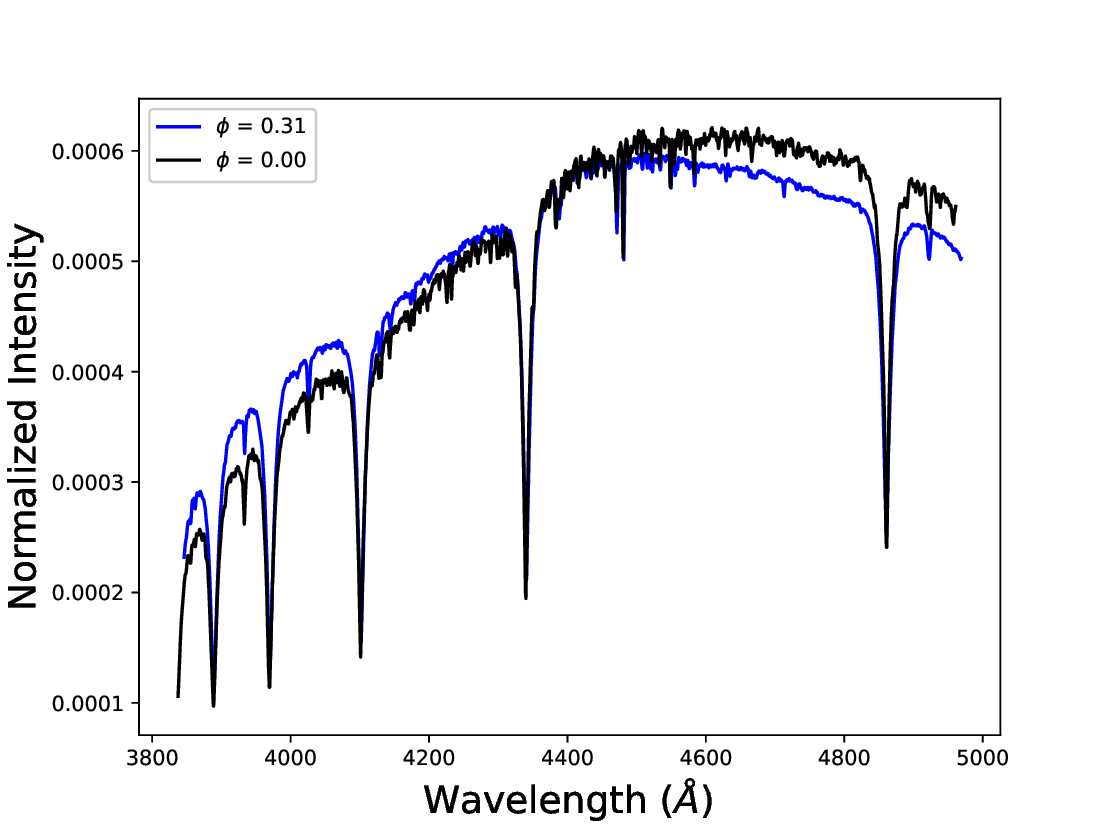}
    \caption{Extracted spectra plotted in ADUs taken at primary eclipse (labeled $\phi = 0.00$) and outside of eclipse (labeled $\phi = 0.31$), normalized.}
    \label{fig:Algol_phi031_phi10_avg_raw}
\end{figure}

Given these constraints, we chose spectra taken near phase angle $\phi = 0.31$\ at an airmass of $\sec z = 1.30$, to subtract from spectra taken at $\phi=0.00$\ ($\sec z = 1.45$). We averaged ten individual spectra taken at each phase to increase the signal-to-noise ratio, so that the subtracted spectrum would show absorption lines into the ultraviolet, where the spectrum is lowest signal-to-noise. Figure~\ref{fig:Algol_phi031_phi10_avg_raw} shows the two spectra, normalized so that they overlap near 4400$\textrm{\AA}$. These are the same spectra that were first shown in Figure~\ref{fig:Algol_phi01_phi10}, but now with their continua. The spectrum from $\phi = 0.31$\ exhibits more light in the ultraviolet than does the spectrum from $\phi = 0.00$. This is expected, since the majority of the UV light from Algol comes from Algol A, which is being blocked during primary eclipse.

Now we must estimate the scale factor by which the $\phi = 0.31$\ spectrum will be multiplied. Our spectral range is fairly close to the bandwidth of the Johnson $B$-band, so we begin by using the $\Delta m_B \approx 1.2$\ that is shown in \citet{Kim1989} for primary eclipse. This magnitude difference equates to a flux ratio $F_{eclipse}/F_{total} \approx 0.33$.

The total flux in the $B$-band is composed of light from stars A, B, and C; {\em i.e.}, $F_{total} = F_A + F_B + F_C$. The observed flux during primary eclipse can be given by $F_{eclipse} \approx 0.33 F_{total} = Y F_A + F_B + F_C$, where $Y$\ is the fraction of the flux still visible from star A, representing how much light from star A remains during primary eclipse. The contribution of light from star A during primary eclipse to the total light during primary eclipse we call our scale factor $SF$ and is written

\begin{equation}
    SF = \frac{Y F_A}{Y F_A + F_B + F_C}
\end{equation}

\noindent which is reminiscent of the algebra used by \citet{Glushneva1968} to find the infrared spectrum of Algol~C. This scale factor SF is applied to one spectrum before subtracting it from another, to properly remove light from Algol A in the spectrum. We assume that the $\phi = 0.31$ spectrum is entirely that of Algol A because that star contributes $\sim 95\%$ of the total light of the system at these wavelengths, and that the $\phi = 0.00$ spectrum is that of Algols A, B, and C combined. This means that the scale factor is applied to the $\phi = 0.31$ spectrum before it is subtracted from the $\phi = 0.00$ spectrum

Determining the scale factor $SF$\ therefore depends on knowing the fractional fluxes of each star in the system. The photometric study by \citet{Kim1989} suggested that the fraction of light from stars A, B, and C in the $B$-band is \{0.854, 0.040,0.106\} respectively, whereas the spectroscopic study of \citet{Kolbas2015} suggested fractions in the same waveband of \{0.943, 0.008, 0.049\}. Various other authors have calculated fractions somewhere between these two sets of values. This provides us with a rough range of possible fractional fluxes. Setting $F_A = 0.85$ on the low end of this range from \citet{Kim1989}, then $Y = 0.21$ (using $F_{eclipse} \approx 0.33 F_{total} = Y F_A + F_B + F_C$ from above), and we determine a value $SF = 0.54$. For $F_A = 0.95$ on the high end of this range from \citet{Kolbas2015}, we compute $Y=0.29$ and $SF = 0.85$. 

\begin{figure}[]
    \centering
    \includegraphics[width = 0.49\textwidth]{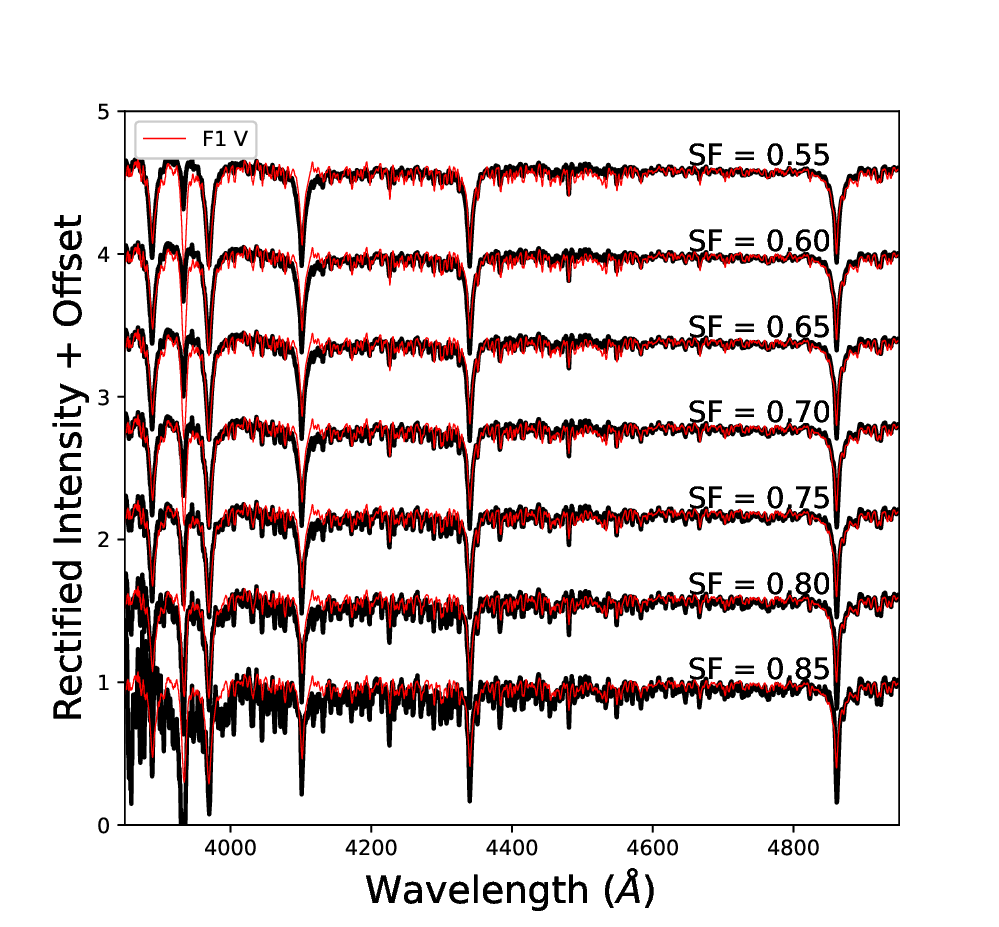}
    \caption{The two spectra shown in Figure~\ref{fig:Algol_phi031_phi10_avg_raw} are subtracted after the $\phi=0.31$ spectrum is scaled by the scale factor SF; the results of subtraction for SF values ranging from 0.55 to 0.85 are shown plotted against an F1V spectral standard.}
    \label{fig:AlgolC_SFvals}
\end{figure}

The choice of the scale factor will have a serious impact on the final interpretation of the spectrum of Algol~C. Figure~\ref{fig:AlgolC_SFvals} shows spectra of Algol C that were computed with scale factors ranging from 0.55 to 0.85, in increments of 0.05, bracketing the most likely set of values as determined in the previous paragraph. Each spectrum is plotted with an F1~V standard star spectrum overplotted for comparison. Changing the scale factor varies several aspects of the spectrum. Most notable is that the strengths of the metal lines are correlated to the scale factor value. The effect of this correlation is most pronounced at shorter wavelengths, and the Ca~{\sc ii}~K line is extremely sensitive. The depth of the Balmer lines is also correlated with scale factor, but this effect is only noticeable from scale factors of about 0.70 and above. The widths of the Balmer line wings is the most noticeable feature in the spectrum that is {\em anti}-correlated with the scale factor because the lines become narrower as the scale factor is increased.

For the purposes of this investigation, it was decided that an iterative approach was best, whereby a scale factor is applied and then the resultant spectrum is compared to standard star spectra. The comparison is then used to inform a change to the scale factor, after which the procedure is performed over again. When comparing the spectrum of Algol~C to the standards, it was further decided that the Balmer lines alone would be used to judge the scale factor choice. This was decided for two reasons. First, the fact that the Balmer line depths and line widths behave oppositely to the scale factor (line depths increase with scale factor while line widths decrease) means that it is possible to find a unique fit to a standard star spectrum using Balmer lines alone. Second, by relying solely on the Balmer lines to determine spectral type we are able to test whether Algol~C is an Am star without inviting circular logic into our result, whereby the metal lines are used to judge whether the scale factor choice is optimized. Such logic would only return a result that we were intending to find from the outset.

\section{Analysis of the Algol C Spectrum}\label{sec:Analysis}

\begin{figure*}[]
    \centering
    \includegraphics[width = 0.99\textwidth]{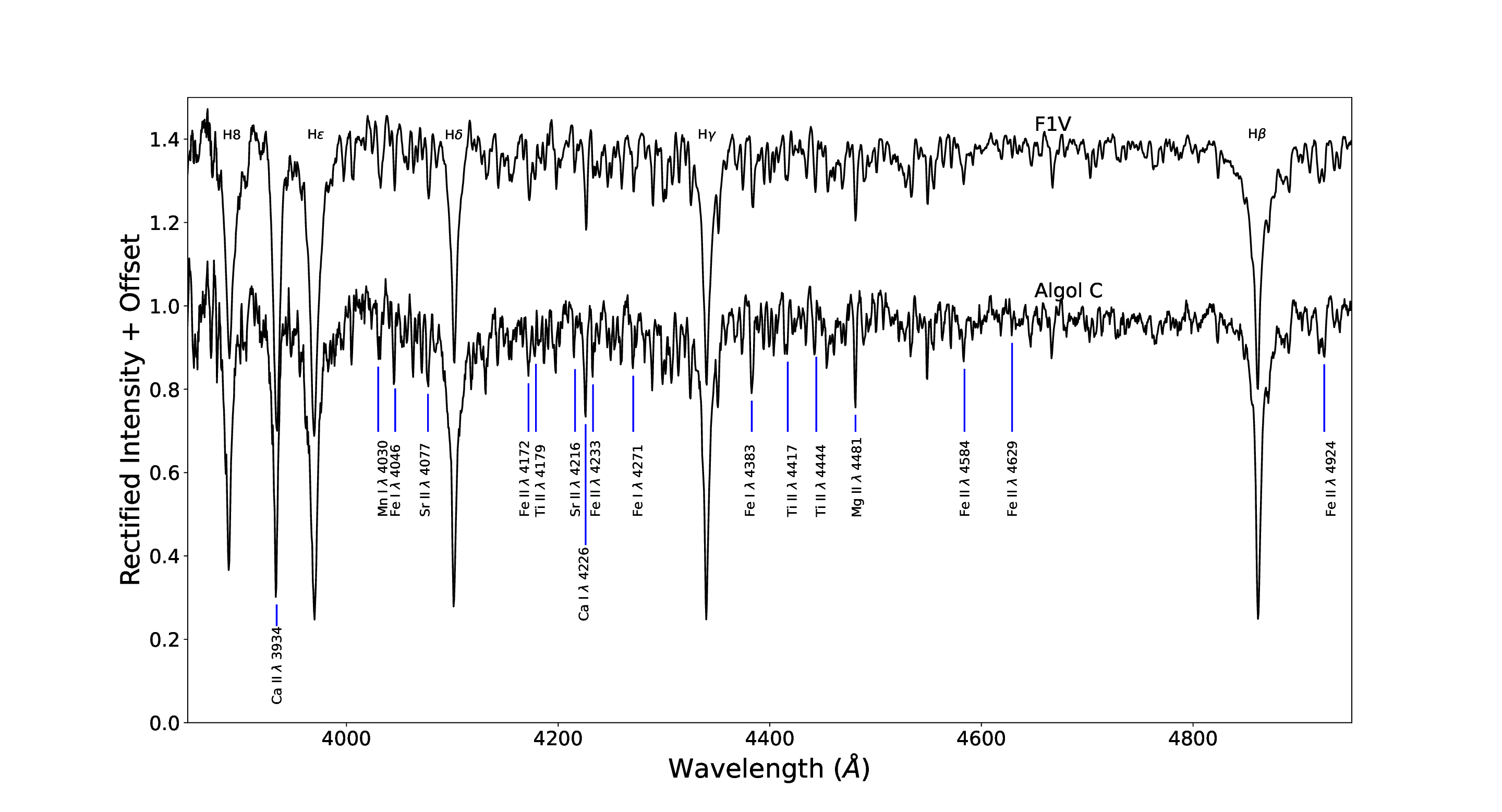}
    \caption{The final spectrum of Algol C (bottom) compared to an F1V comparison spectrum (top). Numerous lines useful in classifying F-type stars are shown.}
    \label{fig:algolC_w_f1v}
\end{figure*}

Figure~\ref{fig:algolC_w_f1v} shows the final spectrum for Algol~C. We used a scale factor of $SF = 0.76$, corresponding to a fraction of the flux still visible from star A during primary eclipse of $Y = 0.27$. 

Notably, we do not see any spectral characteristics that would be considered unique to the spectrum of Algol~B. This includes the strengths of the Cr~{\sc i} $\lambda4254$ and Fe~{\sc i} lines at $\lambda4250$ and $\lambda4260$, as well as the Mn~{\sc i} $\lambda4030$, all of which would be much stronger if Algol~B were contributing substantially to the subtracted spectrum. This was not a surprise: Algol~B is a K2~IV star and, being red, is not expected to contribute much light to the ultraviolet-blue portion of the spectrum, where we are studying. The disentangled spectrum of Algol~B exhibited in \citet{Kolbas2015} shows very weak spectral lines in this wavelength range, so we should not expect much contribution from this source. We ignore Algol~B for the remainder of our analysis.

The metal lines in the spectrum of Algol~C all appear narrow, consistent with its slow rotation rate. The F1V spectral standard, 37~UMa, with a rotational speed of 87 km s$^{-1}$ \citep{Garcia1993}, has similarly narrow lines. Since the resolution of our spectra is $1.4\mathrm{\AA} \approx 100$ km s$^{-1}$, the width of the lines in both spectra is set by the instrument profile without additional Doppler broadening due to rotation. This is helpful in our analysis, because it allows us to directly compare the spectra shown without having to make corrections for line depths due to different stellar rotation speeds.

The spectrum of Algol~C showcases hydrogen Balmer lines that are consistent with an F1V spectral classification. The metallic line spectrum also appears to be consistent with this spectral type -- the strong absorption lines labeled as well as the weaker lines generally -- making allowances for the slightly noisier spectrum of Algol~C. Most notably when looking for characteristics of Am stars, the Ca~{\sc ii}~K and Ca~{\sc i} $\lambda 4226$ lines do not appear weak, and the Sr~{\sc ii} $\lambda 4077$ and $\lambda 4216$ lines are not strong for the metallic-line spectral type.

The evidence from the literature that suggests Algol C is an Am star requires a closer look. \citet{Fletcher1964} claimed that the Ca~{\sc ii} K line was five spectral classes earlier than the Balmer lines, which would be consistent with expectations for Am stars generally. They also claimed that the metal lines were half a spectral type later than were the hydrogen lines, specifically noting that the lines Sr~{\sc ii} $\lambda 4077$, $\lambda 4216$ and Ca~{\sc i} $\lambda 4226$ were all slightly strong. This, however, does not make sense for the spectrum from an Am star. The strength of the Ca~{\sc i} $\lambda 4226$ line is usually weak compared to the metal line spectrum, though not as weak as the Ca~{\sc ii}~K line. The Sr~{\sc ii} lines, however, are usually stronger (later) than even the metal line spectrum \citep{Gray2009}. It would therefore appear that \citet{Fletcher1964} made a mistake. Considering the weak Ca~{\sc ii}~K line in our ``Algol C'' spectrum from Figure~\ref{fig:AlgolC_rect_vs_F0V}, we suspect it is possible that their fault was in the treatment of the spectral continuum.

\citet{Richards1993} showed evidence for a low Calcium abundance in their spectrum of Algol~C. But it also appears that they did not fully trust their abundance analysis, because they classify the star as type F1~V in their Table~1 while referring to it as a ``marginal Am star'' in the text. They label its spectral classification F1~V again in \citet{Richards2012} without further explanation. When we additionally consider that the Calcium abundance determined by \citet{Kolbas2015} was normal, it seems increasingly clear that the current result is the correct one, and that Algol~C is an F1~V star.

We would like to exhaust all possibilities before concluding. Perhaps the spectrum of Algol C has changed since \citeauthor{Fletcher1964}'s observations. This is an intriguing thought, since \citet{Richards1993} showed evidence of a low calcium abundance for Algol~C, but not to the extreme that Fletcher's results implied. Such a finding would be reasonable if Algol~C were gradually transitioning from an Am star to a normal star over the course of the last sixty years.

However, it seems unlikely that the spectrum has changed significantly over the past sixty years; Am stars identified in the early spectroscopic literature remain so today. An example is the prototypical Am star 63~Tau, which was first identified as a chemically peculiar star in \citet{Titus1940} and as an Am star in \citet{Morgan1943}, where the Am class was codified. The Am designation for 63~Tau has remained since, with only minor adjustments \citep[e.g.,][]{Abt1995}.

\section{Conclusion}\label{sec:Conclusion}

We have used a spectral subtraction technique to classify Algol~C. We have observed the Algol system in and out of primary eclipse and, having scaled the spectra to one another appropriately, were able to recover a spectrum for Algol~C without contamination from Algol~A. Our result is that Algol~C has the spectrum of an F1~V star; most notably, we see no sign that it possesses the qualities of a metallic-line star. Our results confirm a recent chemical abundance analysis.


\begin{thebibliography}{}

\bibitem[Abt \& Levy(1985)]{Abt1985} Abt, H.~A. \& Levy, S.~G.\ 1985, \apjs, 59, 229.

\bibitem[Abt \& Morrell(1995)]{Abt1995} Abt, H.~A. \& Morrell, N.~I.\ 1995, \apjs, 99, 135.


\bibitem[Cannon \& Pickering(1918)]{Cannon1918} Cannon, A.~J. \& Pickering, E.~C.\ 1918, Annals of Harvard College Observatory, 91, 1

\bibitem[Charbonneau(1993)]{Charbonneau1993} Charbonneau, P.\ 1993, IAU Colloq. 138: Peculiar versus Normal Phenomena in A-type and Related Stars, 44, 474

\bibitem[Fletcher(1964)]{Fletcher1964} Fletcher, E.~S.\ 1964, \aj, 69, 357. 

\bibitem[Garcia Lopez et al.(1993)]{Garcia1993} Garcia Lopez, R.~J., Rebolo, R., Beckman, J.~E., et al.\ 1993, \aap, 273, 482

\bibitem[Garrison \& Gray(1994)]{Garrison1994} Garrison, R.~F. \& Gray, R.~O.\ 1994, \aj, 107, 1556.

\bibitem[Gebran et al.(2010)]{Gebran2010} Gebran, M., Vick, M., Monier, R., et al.\ 2010, \aap, 523, A71. 

\bibitem[Glushneva \& Esipov(1968)]{Glushneva1968} Glushneva, I.~N. \& Esipov, V.~F.\ 1968, \sovast, 11, 828.




\bibitem[Gray \& Corbally(2009)]{Gray2009} Gray, R.~O. \& Corbally, C.\ Stellar Spectral Classification. Princeton University Press, 2009.

\bibitem[Gray \& Corbally(2014)]{Gray2014} Gray, R.~O. \& Corbally, C.~J.\ 2014, \aj, 147, 80.

\bibitem[Griffin \& Griffin(1986)]{Griffin1986} Griffin, R. \& Griffin, R.\ 1986, Journal of Astrophysics and Astronomy, 7, 195. 

\bibitem[Hadrava(1995)]{Hadrava1995} Hadrava, P.\ 1995, \aaps, 114, 393.

\bibitem[Hall(1939)]{Hall1939} Hall, J.~S.\ 1939, \apj, 90, 449.


\bibitem[Kim(1989)]{Kim1989} Kim, H.-I.\ 1989, \apj, 342, 1061.

\bibitem[Kolbas et al.(2015)]{Kolbas2015} Kolbas, V., Pavlovski, K., Southworth, J., et al.\ 2015, \mnras, 451, 4150.

\bibitem[Meltzer(1957)]{Meltzer1957} Meltzer, A.~S.\ 1957, \apj, 125, 359. 


\bibitem[Morgan et al.(1943)]{Morgan1943} Morgan, W.~W., Keenan, P.~C., \& Kellman, E.\ 1943, Chicago, Ill., The University of Chicago press.

\bibitem[Richards et al.(1987)]{Richards1987} Richards, M.~T., Bolton, C.~T., \& Mochnacki, S.~W.\ 1987, \baas

\bibitem[Richards et al.(1988)]{Richards1988} Richards, M.~T., Mochnacki, S.~W., \& Bolton, C.~T.\ 1988, \aj, 96, 326. 


\bibitem[Richards(1993)]{Richards1993} Richards, M.~T.\ 1993, \apjs, 86, 255. 

\bibitem[Richards et al.(2012)]{Richards2012} Richards, M.~T., Agafonov, M.~I., \& Sharova, O.~I.\ 2012, \apj, 760, 8.


\bibitem[Struve \& Sahade(1957)]{Struve1957} Struve, O. \& Sahade, J.\ 1957, \pasp, 69, 41.

\bibitem[Titus \& Morgan(1940)]{Titus1940} Titus, J. \& Morgan, W.~W.\ 1940, \apj, 92, 256.

\bibitem[Zavala et al.(2010)]{Zavala2010} Zavala, R.~T., Hummel, C.~A., Boboltz, D.~A., et al.\ 2010, \apjl, 715, L44.

\end{thebibliography}
\end{document}